\documentclass{article}

     \usepackage[nonatbib,preprint]{neurips_2020}

\usepackage[utf8]{inputenc} 
\usepackage[T1]{fontenc}    
\usepackage{hyperref}       
\usepackage{url}            
\usepackage{booktabs}       
\usepackage{amsfonts}       
\usepackage{nicefrac}       
\usepackage{microtype}      
\usepackage{slashbox}
\usepackage{amsmath}
\usepackage{ dsfont }
\usepackage{graphicx}
\usepackage{subcaption}
\usepackage[backend=biber,style=science,sorting=none]{biblatex}
\usepackage{authblk}
\usepackage{multirow}
\usepackage{float}

\title{Leveraging Protein Language Model Embeddings for Catalytic Turnover Prediction of Adenylate Kinase Orthologs in a Low-Data Regime}

\author[1,2,{}\textdagger]{\textbf{Duncan F. Muir}}
\author[1,3,{}\textdagger]{\textbf{Parker Grosjean}}
\author[1,4]{\textbf{Margaux M. Pinney}}
\author[3,5]{\textbf{Michael J. Keiser}}

\affil[1]{\footnotesize Department of Biochemistry and Biophysics, University of California, San Francisco, San Francisco, CA, USA}
\affil[2]{\footnotesize Program in Biophysics, University of California, San Francisco, San Francisco, CA, USA}
\affil[3]{\footnotesize Institute for Neurodegenerative Diseases, University of California, San Francisco, San Francisco, CA, USA}
\affil[4]{\footnotesize Valhalla Fellow, University of California San Francisco, San Francisco, CA, USA}
\affil[5]{\footnotesize Department of Pharmaceutical Chemistry, University of California, San Francisco, San Francisco, CA, USA}

\affil[{}\textdagger]{\footnotesize These authors contributed equally. \{dmuir, pgrosjean\}@keiserlab.org}

\begin{document}

\maketitle

\begin{abstract}
Accurate prediction of enzymatic activity from amino acid sequences could drastically accelerate enzyme engineering for applications such as bioremediation and therapeutics development. In recent years, Protein Language Model (PLM) embeddings have been increasingly leveraged as the input into sequence-to-function models. Here, we use consistently collected catalytic turnover observations for 175 orthologs of the enzyme Adenylate Kinase (ADK) as a test case to assess the use of PLMs and their embeddings in enzyme kinetic prediction tasks. In this study, we show that nonlinear probing of PLM embeddings outperforms baseline embeddings (one-hot-encoding) and the specialized $k_{cat}$ (catalytic turnover number) prediction models DLKcat and CatPred.
We also compared fixed and learnable aggregation of PLM embeddings for $k_{cat}$ prediction and found that transformer-based learnable aggregation of amino-acid PLM embeddings is generally the most performant. Additionally, we found that ESMC 600M embeddings marginally outperform other PLM embeddings for $k_{cat}$ prediction. We explored Low-Rank Adaptation (LoRA) masked language model fine-tuning and direct fine-tuning for sequence-to-$k_{cat}$ mapping, where we found no difference or a drop in performance compared to zero-shot embeddings, respectively. And we investigated the distinct hidden representations in PLM encoders and found that earlier layer embeddings perform comparable to or worse than the final layer. Overall, this study assesses the state of the field for leveraging PLMs for sequence-to-$k_{cat}$ prediction on a set of diverse ADK orthologs.\footnote{Code available at https://github.com/keiserlab/face-plm.}
\end{abstract}

\section{Introduction}
\label{intro}

Quantitative prediction of enzymatic activity from amino acid sequence alone holds great promise to advance the field of enzyme design and engineering. Enzymes with improved activity have numerous downstream applications, ranging from bioremediation (i.e., degradation of post-consumer plastic) to therapeutics (i.e., enzyme replacement therapies). Historically, optimization of enzyme activity has been performed through directed evolution \cite{Arnold1996-mc}. In recent years this process has been augmented via machine learning – learning a sequence-to-function model – to direct the pseudo-evolutionary processes \cite{Yang2019-nr, Yang2024-tq, Li2024-to, Jiang2024-zk}. Recent prediction methods have used embeddings generated from Protein Language Models (PLMs) as model inputs \cite{Jiang2024-zk}. PLMs are thought to reflect the ‘fitness landscape’ of protein sequences by learning the evolutionary conservation of amino acids contextualized by the rest of the protein sequence. The field has largely turned to PLMs under the hypothesis that these embeddings are meaningful representations for learning sequence-to-function mappings \cite{Li2022-wr, Boorla2025-uk}.

However, to our knowledge, there has been no systematic benchmarking of PLM methods for sequence-to-function mapping of enzymatic activity prediction. To address this gap, we focused on sequence-to-$k_{cat}$ prediction with a unique, self-consistent dataset of 175 $k_{cat}$ values for orthologs of a single enzyme, ADK \cite{Muir2024-ph}. While traditional machine-learning-guided directed evolution relies on predicting the effect of low edit distance variants from a starting wild-type sequence, this dataset encompasses a broader sequence space of diverse ADK orthologs (42\% average pairwise sequence identity) \cite{Muir2024-ph}. This enables a new benchmarking task for testing sequence-to-function modeling in a larger, more diverse sequence space. In this study, we leveraged six state-of-the-art pre-trained PLMs and investigated fixed and learnable aggregation methods, LoRA fine-tuning to augment PLM embeddings, and direct LoRA fine-tuning for sequence-to-$k_{cat}$ prediction. In total, we assessed the state of the field for sequence-to-function prediction leveraging PLMs in a low-data regime.

\begin{figure}[h]
    \centering
    \includegraphics[width=1\linewidth]{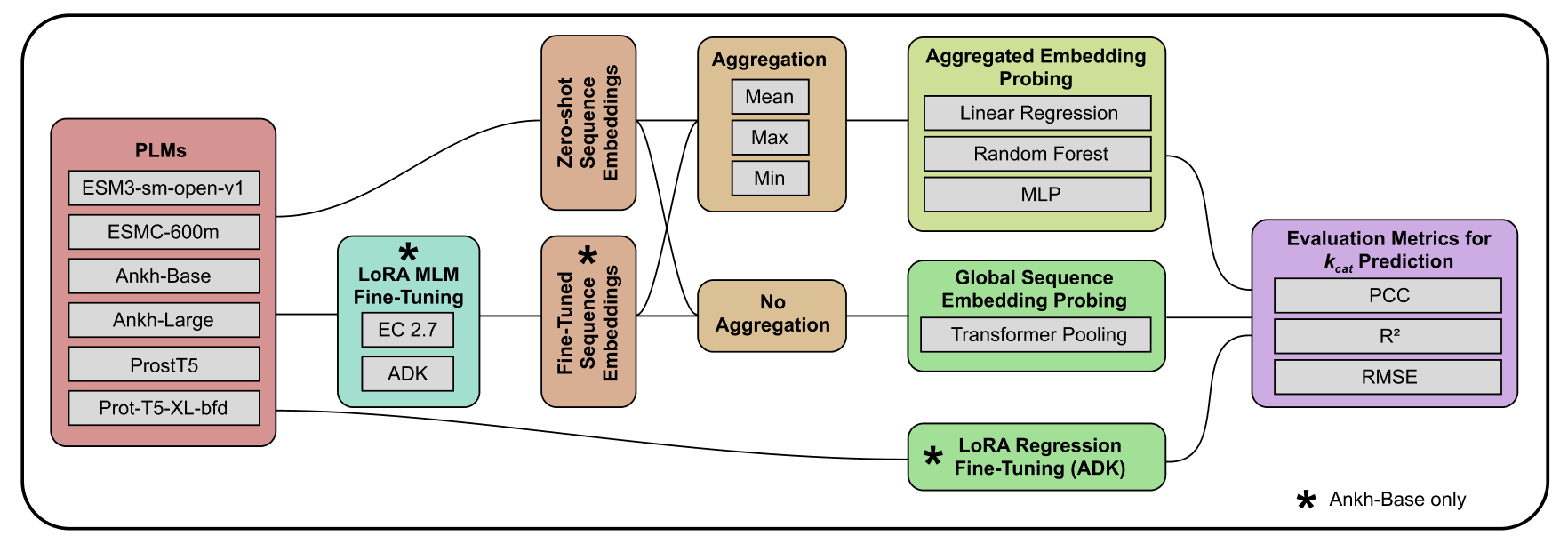}
    \caption{Overview of the experiments and benchmarking performed in this study}
    \label{fig:figure1_corrected}
\end{figure}

\section{Background}

\subsection{Dataset}

In this study we investigate the task of predicting the catalytic turnover number ($k_{cat}$) of amino-acid sequences for 175 orthologs of the enzyme Adenylate Kinase (ADK) from a recently released dataset collected under consistent experimental conditions \cite{Muir2024-ph}. A unique feature of ADK activity is that the opening of the LID domain is rate-limiting for catalysis \cite{Wolf-Watz2004-nf}, and different architectures or “lid types” of ADK have significantly different activity \cite{Muir2024-ph}. To minimize potential shortcut learning or memorization of distributional differences by lid type, we used a 5-fold lid-aware train:test cross-validation strategy when training and evaluating all regression tasks (\textbf{Appendix A}). As in other $k_{cat}$ prediction studies, we $log_{10}$ normalize $k_{cat}$ values for training and evaluation.

\subsection{Protein Language Models}

We assessed six pre-trained protein language models (PLMs): Ankh-base \cite{Elnaggar2023-dw}, Ankh-large \cite{Elnaggar2023-dw}, Prot-T5-XL-BFD \cite{Elnaggar2022-ka}, ProstT5 \cite{Heinzinger2024-bi}, ESMC-600M \cite{ESM-Team2024-qb}, and ESM3-open \cite{Hayes2025-wh}. These protein language models all leverage masked language modeling objectives for pre-training. Some of the models, including ESM3 and ProstT5, explicitly train on protein structure data to relate sequence with structure. We focused on embeddings extracted from the encoders of each model.

\subsection{Evaluation Metrics}

We evaluated all models for the $log_{10}(k_{cat})$ regression task. We used the Pearson Correlation Coefficient (PCC), root mean square error (RMSE), and the Coefficient of determination $(R^{2})$ as performance metrics. We additionally used the RankMe metric \cite{Garrido2022-fb} to quantify the effective rank of the models' embeddings to assess how embedding rank affects downstream performance. We report the metric means plus or minus the standard deviation across the five-fold cross-validation split.

\section{Experiments}
We evaluated the effects of different aggregation methods, probing techniques, and fine-tuning methods on sequence to $k_{cat}$ prediction leveraging PLMs and their embeddings. We additionally compared the optimal PLM-based sequence to $k_{cat}$ probing method against a set of baseline models. Finally, we investigated whether intermediate layer embeddings offer better representations than the last layer embeddings for the prediction of $k_{cat}$.

\subsection{Experiment 1: Fixed aggregation probing of zero-shot PLM embeddings for ADK $k_{cat}$ Prediction}
Sequence-to-function models require an encoding method that takes in different length sequences and generates a single output. Current strategies leveraging PLM embeddings largely rely on small lightweight models built atop a protein sequence's mean aggregated amino acid embeddings. In this experiment, we extracted zero-shot embeddings from the final layer of all six pre-trained PLMs described in \textbf{Section 2.2}. We used mean, max, and min pooling to generate a single embedding per sequence. We then performed linear, random forest, and multi-layer perceptron (MLP) probing for $k_{cat}$ evaluated by the metrics described in \textbf{Section 2.3} on the five-fold cross-validation split described in \textbf{Section 2.1}.

\subsection{Experiment 2: Learnable aggregation probing of zero-shot PLM Embeddings for ADK $k_{cat}$ Prediction}
Here, we sought to assess a learnable aggregation function for predicting $k_{cat}$ directly form the amino acid embeddings. We tested both a lightweight single-head attention layer and a larger Transformer encoder (ViT-Tiny architecture) trained directly on amino-acid PLM embeddings padded to a sequence length of 245 (the longest sequence in the dataset). The models were trained for 1000 epochs with a linear warm-up of 50 epochs, a base learning rate of 1e-6, and a cosine decay to 1e-8. These models were evaluated following the approach outlined in \textbf{Section 3.1}.

\subsection{Experiment 3: Masked Language Model LoRA fine-tuning and probing for ADK $k_{cat}$ prediction}

We next sought to determine if parameter efficient masked language model (MLM) fine-tuning on protein subspaces closer to the ADK orthologs in the dataset is beneficial for probing on the $k_{cat}$ regression task. Here we investigated two levels of protein subspaces.

The first dataset was generated by downloading all enzymes that transfer phosphorus-containing groups (EC 2.7) and filtering for the sequences from the organisms in the ADK ortholog dataset (n=119781). This dataset represents an enzyme sequence subspace with a broadly similar function to ADK.

The second dataset was generated by downloading all known orthologs of ADK from UniProt across all species. This dataset represents all orthologs of ADK (n=48794).

We then performed parameter-efficient MLM fine-tuning of the Ankh-base model with LoRA for 1000 epochs. Ankh-base was chosen for fine-tuning, as it has $\sim$740M parameters and is easily fine-tuned on GPUs containing as little as 24 GB of memory. We used a linear warmup of 10 epochs, a maximum learning rate of 1e-3 followed by a cosine decay to 1e-6. To determine optimal LoRA hyperparameters, we performed a sweep for the LoRA rank and the LoRA alpha parameter. We determined that a rank of 16 and an alpha of 32 was the most parameter-efficient combination as evaluated via the cross-entropy MLM loss. Amino acid embeddings were extracted from the two fine-tuned PLMs (EC 2.7, ADK) and probing for $k_{cat}$ was performed and evaluated using the metrics defined in \textbf{Section 2.3}.

We also calculated the RankMe scores to quantify the role of fine-tuning on amino acid embedding rank and its effect on probing performance. A RankMe score was calculated for both fine-tuned models using the amino acid embeddings from all sequences in the ADK ortholog dataset. The RankMe scores were calculated as defined in \cite{Garrido2022-fb}.

\subsection{Experiment 4: Direct PLM LoRA fine-tuning for ADK $k_{cat}$ prediction}
We next evaluated direct parameter efficient fine-tuning of the Ankh-base model for $k_{cat}$ prediction. To perform the direct LoRA fine tuning we concatenated a CLS token to all input amino acid embeddings and padded to a sequence length of 512. The CLS token was used as input into a single linear layer for the regression task with a mean squared error loss function. We use the same LoRA hyperparameters as described in \textbf{Section 3.3} with the exception of an increased LoRA dropout (0.5). This model was trained for 1000 epochs on each of the five train splits and evaluated using the metrics defined in \textbf{Section 2.3}.

\subsection{Experiment 5: Comparison to general $k_{cat}$ prediction baselines}
We evaluated pre-trained models for direct $k_{cat}$ prediction for comparison to our PLM probing methods. These models were trained on pairs of substrate SMILES strings and amino acid sequence embeddings to predict enzyme kinetic parameters. Here we evaluated DLKcat \cite{Li2022-wr} and CatPred \cite{Boorla2025-uk} as machine learning model baselines for direct $k_{cat}$ prediction. Both models were used in inference with ADP as the substrate input and the ADK sequences as the sequence input.

\subsection{Experiment 6: Probing embeddings from all PLM hidden layers}
We next evaluated whether distinct hidden layer embeddings from PLMs have differential performance as input for the $k_{cat}$ regression task. Here we leveraged the Ankh-base PLM and extracted the hidden layer embeddings for all 49 layers. We then trained learnable aggregation models, and mean aggregation random forest models for the $k_{cat}$ prediction.

\section{Results}

\subsection{ESMC 600m embeddings exhibit top performance in fixed aggregation probing}

The results of the 5-fold cross-validation probing with a linear regression model, a random forest model, and a 2-layer multi-layer perceptron (MLP) model are shown in \textbf{Table 1}. We omit the results of the max pooling from \textbf{Table 1} as they never performed better than mean or min pooling across all the models and probing methods. These results suggest different aggregation methods perform better for different PLM embeddings and that mean aggregation does not always generate the best representation for downstream regression tasks. In fact, the best-performing model we observed was on min pooled embeddings from the ESMC 600M model. On average, MLP probing was the most performant across the 5-fold train:test splits, outperforming linear and random forest probing. However, many of these evaluation metric scores are close and likely suggest only marginal gains in performance. In an outlier, ESM3 embeddings yielded poor performance, which we believe arises from overfitting to the train sets due to large values in the embedding vectors. We believe this is the case because we observe larger L2 norms for the ESM3 embeddings compared to other PLMs (\textbf{Appendix B}). We additionally performed a replicate training exercise with MLP probing to confirm consistent performance across multiple random seeds (\textbf{Appendix C}).

\subsection{Learnable aggregation provides marginal improvement over fixed aggregation for $k_{cat}$ prediction}

The results of the learnable aggregation probing are shown in \textbf{Table 2}. For all models except ESMC 600M, we observed a marginal improvement in the RMSE compared to the fixed aggregation probing.

\begin{table}[H]
\caption{Comparing fixed aggregation and probing methods on PLM embeddings for $k_{cat}$ regression}
\centering

\begin{tabular}{llllll}
\toprule
  PLM Model & Aggr. & Probe & RMSE ($\downarrow$) & PCC ($\uparrow$) & $R^2$ ($\uparrow$) \\
\midrule
\multirow[t]{6}{*}{Ankh-base} & \multirow[t]{3}{*}{mean} & Lin Reg & 0.907 $\pm$ 0.353 & 0.354 $\pm$ 0.269 & -1.445 $\pm$ 1.924 \\
 &  & MLP & 0.504 $\pm$ 0.051 & 0.570 $\pm$ 0.149 & 0.326 $\pm$ 0.150 \\
 &  & R.F. & 0.517 $\pm$ 0.040 & 0.548 $\pm$ 0.150 & 0.289 $\pm$ 0.156 \\
\cline{2-6}
 & \multirow[t]{3}{*}{min} & Lin Reg & 0.621 $\pm$ 0.047 & 0.420 $\pm$ 0.132 & -0.035 $\pm$ 0.254 \\
 &  & MLP & 0.520 $\pm$ 0.044 & 0.541 $\pm$ 0.138 & 0.284 $\pm$ 0.143 \\
 &  & R.F. & 0.564 $\pm$ 0.077 & 0.443 $\pm$ 0.148 & 0.170 $\pm$ 0.136 \\
\cline{1-6} \cline{2-6}
\multirow[t]{6}{*}{Ankh-large} & \multirow[t]{3}{*}{mean} & Lin Reg & 0.682 $\pm$ 0.054 & 0.402 $\pm$ 0.176 & -0.271 $\pm$ 0.413 \\
 &  & MLP & 0.511 $\pm$ 0.051 & 0.562 $\pm$ 0.106 & 0.314 $\pm$ 0.113 \\
 &  & R.F. & 0.524 $\pm$ 0.054 & 0.551 $\pm$ 0.088 & 0.281 $\pm$ 0.089 \\
\cline{2-6}
 & \multirow[t]{3}{*}{min} & Lin Reg & 0.578 $\pm$ 0.073 & 0.412 $\pm$ 0.113 & 0.126 $\pm$ 0.131 \\
 &  & MLP & 0.549 $\pm$ 0.073 & 0.483 $\pm$ 0.054 & 0.220 $\pm$ 0.056 \\
 &  & R.F. & 0.586 $\pm$ 0.083 & 0.390 $\pm$ 0.089 & 0.113 $\pm$ 0.072 \\
\cline{1-6} \cline{2-6}
\multirow[t]{6}{*}{ESM3 Open V1} & \multirow[t]{3}{*}{mean} & Lin Reg & 0.666 $\pm$ 0.059 & 0.427 $\pm$ 0.184 & -0.199 $\pm$ 0.392 \\
 &  & MLP & 0.763 $\pm$ 0.105 & 0.434 $\pm$ 0.179 & -0.622 $\pm$ 0.649 \\
 &  & R.F. & 0.507 $\pm$ 0.054 & 0.583 $\pm$ 0.095 & 0.327 $\pm$ 0.095 \\
\cline{2-6}
 & \multirow[t]{3}{*}{min} & Lin Reg & 0.569 $\pm$ 0.045 & 0.464 $\pm$ 0.104 & 0.145 $\pm$ 0.153 \\
 &  & MLP & 3.528 $\pm$ 0.846 & 0.022 $\pm$ 0.137 & -32.459 $\pm$ 12.702 \\
 &  & R.F. & 0.562 $\pm$ 0.071 & 0.469 $\pm$ 0.149 & 0.176 $\pm$ 0.115 \\
\cline{1-6} \cline{2-6}
\multirow[t]{6}{*}{ESMC 600m} & \multirow[t]{3}{*}{mean} & Lin Reg & 0.807 $\pm$ 0.068 & 0.334 $\pm$ 0.117 & -0.740 $\pm$ 0.430 \\
 &  & MLP & 0.501 $\pm$ 0.072 & 0.586 $\pm$ 0.101 & 0.346 $\pm$ 0.115 \\
 &  & R.F. & 0.515 $\pm$ 0.073 & 0.568 $\pm$ 0.058 & 0.313 $\pm$ 0.071 \\
\cline{2-6}
 & \multirow[t]{3}{*}{min} & Lin Reg & 0.528 $\pm$ 0.063 & 0.573 $\pm$ 0.050 & 0.271 $\pm$ 0.096 \\
 &  & MLP & \textbf{0.473 $\pm$ 0.069} & \textbf{0.647 $\pm$ 0.078} & \textbf{0.416 $\pm$ 0.110} \\
 &  & R.F. & 0.547 $\pm$ 0.057 & 0.505 $\pm$ 0.075 & 0.220 $\pm$ 0.060 \\
\cline{1-6} \cline{2-6}
\multirow[t]{6}{*}{ProstT5} & \multirow[t]{3}{*}{mean} & Lin Reg & 0.719 $\pm$ 0.058 & 0.348 $\pm$ 0.114 & -0.391 $\pm$ 0.365 \\
 &  & MLP & 0.520 $\pm$ 0.046 & 0.529 $\pm$ 0.170 & 0.280 $\pm$ 0.160 \\
 &  & R.F. & 0.546 $\pm$ 0.062 & 0.493 $\pm$ 0.086 & 0.222 $\pm$ 0.081 \\
\cline{2-6}
 & \multirow[t]{3}{*}{min} & Lin Reg & 0.656 $\pm$ 0.083 & 0.323 $\pm$ 0.170 & -0.139 $\pm$ 0.286 \\
 &  & MLP & 0.561 $\pm$ 0.068 & 0.454 $\pm$ 0.076 & 0.183 $\pm$ 0.071 \\
 &  & R.F. & 0.567 $\pm$ 0.075 & 0.441 $\pm$ 0.059 & 0.167 $\pm$ 0.053 \\
\cline{1-6} \cline{2-6}
\multirow[t]{6}{*}{ProtT5-XL-BFD} & \multirow[t]{3}{*}{mean} & Lin Reg & 0.766 $\pm$ 0.113 & 0.388 $\pm$ 0.195 & -0.557 $\pm$ 0.426 \\
 &  & MLP & 0.502 $\pm$ 0.046 & 0.568 $\pm$ 0.149 & 0.329 $\pm$ 0.162 \\
 &  & R.F. & 0.525 $\pm$ 0.055 & 0.541 $\pm$ 0.123 & 0.275 $\pm$ 0.121 \\
\cline{2-6}
 & \multirow[t]{3}{*}{min} & Lin Reg & 0.572 $\pm$ 0.065 & 0.476 $\pm$ 0.188 & 0.100 $\pm$ 0.336 \\
 &  & MLP & 0.496 $\pm$ 0.047 & 0.576 $\pm$ 0.153 & 0.343 $\pm$ 0.169 \\
 &  & R.F. & 0.552 $\pm$ 0.062 & 0.506 $\pm$ 0.076 & 0.209 $\pm$ 0.047 \\
\cline{1-6} \cline{2-6}
\bottomrule
\end{tabular}
\end{table}

\begin{table}[h]
\caption{Comparing Learnable Pooling with Transformer vs. MLP on Aggregated Embeddings}
\centering
\begin{tabular}{lllll}
\toprule
 PLM Model & Aggregation    & RMSE ($\downarrow$) & PCC ($\uparrow$) & $R^2$ ($\uparrow$) \\

\midrule
\multirow[t]{2}{*}{Ankh-base} & Agg. (mean) & 0.504 $\pm$ 0.051 & 0.570 $\pm$ 0.149 & 0.326 $\pm$ 0.150 \\
 & Learned & \textbf{0.493 $\pm$ 0.045} & \textbf{0.575 $\pm$ 0.166} & \textbf{0.342 $\pm$ 0.203} \\
\cline{1-5}
\multirow[t]{2}{*}{Ankh-large} & Agg. (mean) & 0.511 $\pm$ 0.051 & 0.562 $\pm$ 0.106 & 0.314 $\pm$ 0.113 \\
 & Learned & \textbf{0.483 $\pm$ 0.056} & \textbf{0.619 $\pm$ 0.122} & \textbf{0.380 $\pm$ 0.152} \\
\cline{1-5}
\multirow[t]{2}{*}{ESM3 Open V1} & Agg. (mean) & 0.763 $\pm$ 0.105 & 0.434 $\pm$ 0.179 & -0.622 $\pm$ 0.649 \\
 & Learned & \textbf{0.542 $\pm$ 0.064} & \textbf{0.499 $\pm$ 0.070} & \textbf{0.235 $\pm$ 0.086} \\
\cline{1-5}
\multirow[t]{2}{*}{ESMC 600M} & Agg. (min) & \textbf{0.473 $\pm$ 0.069} & \textbf{0.647 $\pm$ 0.078} & \textbf{0.416 $\pm$ 0.110} \\
 & Learned & 0.489 $\pm$ 0.066 & 0.608 $\pm$ 0.106 & 0.373 $\pm$ 0.121 \\
\cline{1-5}
\multirow[t]{2}{*}{ProstT5} & Agg. (mean) & 0.520 $\pm$ 0.046 & 0.529 $\pm$ 0.170 & 0.280 $\pm$ 0.160 \\
 & Learned & \textbf{0.497 $\pm$ 0.057} & \textbf{0.594 $\pm$ 0.128} & \textbf{0.341 $\pm$ 0.161} \\
\cline{1-5}
\multirow[t]{2}{*}{ProtT5-XL-BFD} & Agg. (min) & 0.496 $\pm$ 0.047 & 0.576 $\pm$ 0.153 & 0.343 $\pm$ 0.169 \\
 & Learned & \textbf{0.493 $\pm$ 0.053} & \textbf{0.580 $\pm$ 0.175} & \textbf{0.349 $\pm$ 0.192} \\
\cline{1-5}
\bottomrule
\end{tabular}
\end{table}

\subsection{Masked Language Model LoRA fine-tuning does not consistently improve prediction performance.}

To investigate whether masked language model PLM fine-tuning could provide more performant embeddings for $k_{cat}$ prediction, we fine-tuned on two protein subspaces (EC2.7, and ADK homologs). After fine-tuning on protein subspaces closer to the ADK orthologs, we observed a small increase in the RankMe score, particularly on the ADK dataset (\textbf{Table 3}). This suggests that MLM fine-tuning on protein subspaces does marginally increase embedding rank. However, the relative increases in the embedding rank were small and suggested that the original embedding space did not suffer from rank collapse. We compared the regression task performance using the learnable aggregation over the zero-shot embeddings vs the fine-tuned model embeddings. Although fine-tuning with an MLM objective increased the effective rank of the amino acid embeddings, it did not meaningfully change downstream $k_{cat}$ prediction performance.

\begin{table}[h]
\caption{Comparing MLM LoRA fine-tuning of Ankh-base model on protein subspaces versus zero-shot embeddings for $k_{cat}$ prediction}
\centering
\begin{tabular}{llllll}
\toprule
 & Fine-tune Data & RMSE ($\downarrow$) & PCC ($\uparrow$) & $R^2$ ($\uparrow$) & RankMe ($\uparrow$) \\
 \midrule
 & ADK            & 0.509 $\pm$ 0.038 & 0.546 $\pm$ 0.146 & 0.308 $\pm$ 0.157  &     \textbf{487.09 }     \\
  \cline{1-6}
 & EC 2.7         & 0.500 $\pm$ 0.035 & \textbf{0.582 $\pm$ 0.141} & 0.332 $\pm$ 0.159    & 483.85        \\
  \cline{1-6}
 & None           & \textbf{0.493 $\pm$ 0.045} & 0.575 $\pm$ 0.166 & \textbf{0.342 $\pm$ 0.203} & 475.79 \\        
\bottomrule
\end{tabular}
\end{table}

\subsection{Direct LoRA fine-tuning decreases performance.}
Direct fine-tuning of the Ankh-base led to a substantial performance decrease under five-fold cross-validation testing (\textbf{Table 4}). The performance disparity on the train versus test sets suggests the model quickly over-fits the training data. This was slightly remedied by regularization. A larger LoRA dropout increased the downstream regression performance, but it never reached the accuracy of nonlinear probing of zero-shot PLM embeddings.

\begin{table}[h]
\caption{Comparing direct LoRA fine-tuning on the $k_{cat}$ prediction task}
\centering
\begin{tabular}{lllll}
\toprule
 & Regression Method & RMSE ($\downarrow$) & PCC ($\uparrow$) & $R^2$ ($\uparrow$)  \\
 \midrule
 &  Ankh-Base Direct LoRA Fine Tune   &  0.605 $\pm$ 0.109 & 0.489 $\pm$ 0.106 & 0.037 $\pm$ 0.243    \\
  \cline{1-5}
 &  Ankh-base Zero-Shot Learned Agg.    & \textbf{0.493 $\pm$ 0.045} & \textbf{0.575 $\pm$ 0.166} & \textbf{0.342 $\pm$ 0.203} \\        
\bottomrule
\end{tabular}
\end{table}

\subsection{Top-performing PLM probing outperform prior specialized $k_{cat}$ prediction models}
 This study's top-performing models outperformed both specialized direct $k_{cat}$ prediction models: DLKcat and CatPred. This suggests that training on large databases with less sequence coverage per protein (BRENDA contains 9 ADK sequences with $k_{cat}$ values) does not enable meaningful prediction of $k_{cat}$ among members of the ADK family (\textbf{Table 5}). Additionally, our PLM probing outperforms traditional encoding techniques such as BLOSUM62 and one-hot-encoding (\textbf{Appendix D}).

\begin{table}[h]
\caption{Comparing against public $k_{cat}$ prediction models}
\centering
\begin{tabular}{lllll}
\toprule
 & Regression Method & RMSE ($\downarrow$) & PCC ($\uparrow$) & $R^2$ ($\uparrow$)  \\
 \midrule
 &  CatPred   &  1.443 $\pm$ 0.022 & 0.288 $\pm$ 0.128 & -5.527 $\pm$ 3.008    \\
  \cline{1-5}
 &  DLKcat   &  1.873 $\pm$ 0.078 & -0.093 $\pm$ 0.044 & -8.408 $\pm$ 2.189    \\
  \cline{1-5}
 &  ESMC 600M (Min Agg. MLP) & \textbf{0.473 $\pm$ 0.069} & \textbf{0.647 $\pm$ 0.078} & \textbf{0.416 $\pm$ 0.110} \\   
  \cline{1-5}
 &  Ankh-large (Learnable Agg.)    & 0.483 $\pm$ 0.056 & 0.619 $\pm$ 0.122 & 0.380 $\pm$ 0.152 \\  
\bottomrule
\end{tabular}
\end{table}

\subsection{Intermediate PLM layer embeddings provide no performance boost compared to the final layer embeddings for Ankh-base}

We found that PLM amino acid embeddings change their magnitudes as measured by L2 norm throughout the layers, with an abrupt change at the last layer (\textbf{Figure 2A}). To assess comparative regression performance with a learnable aggregation probing method, we applied layer norm to each amino acid embedding for layers 31 to 47 to ensure numerical stability during training. Without layer norm, the embeddings are large and cause exploding gradients, which causes model training to fail. We found that probing these layers was at best comparable to probing the final layer embeddings (layer 48, \textbf{Figure 2B}). 

\begin{figure}[h]
    \centering
    \includegraphics[width=1\linewidth]{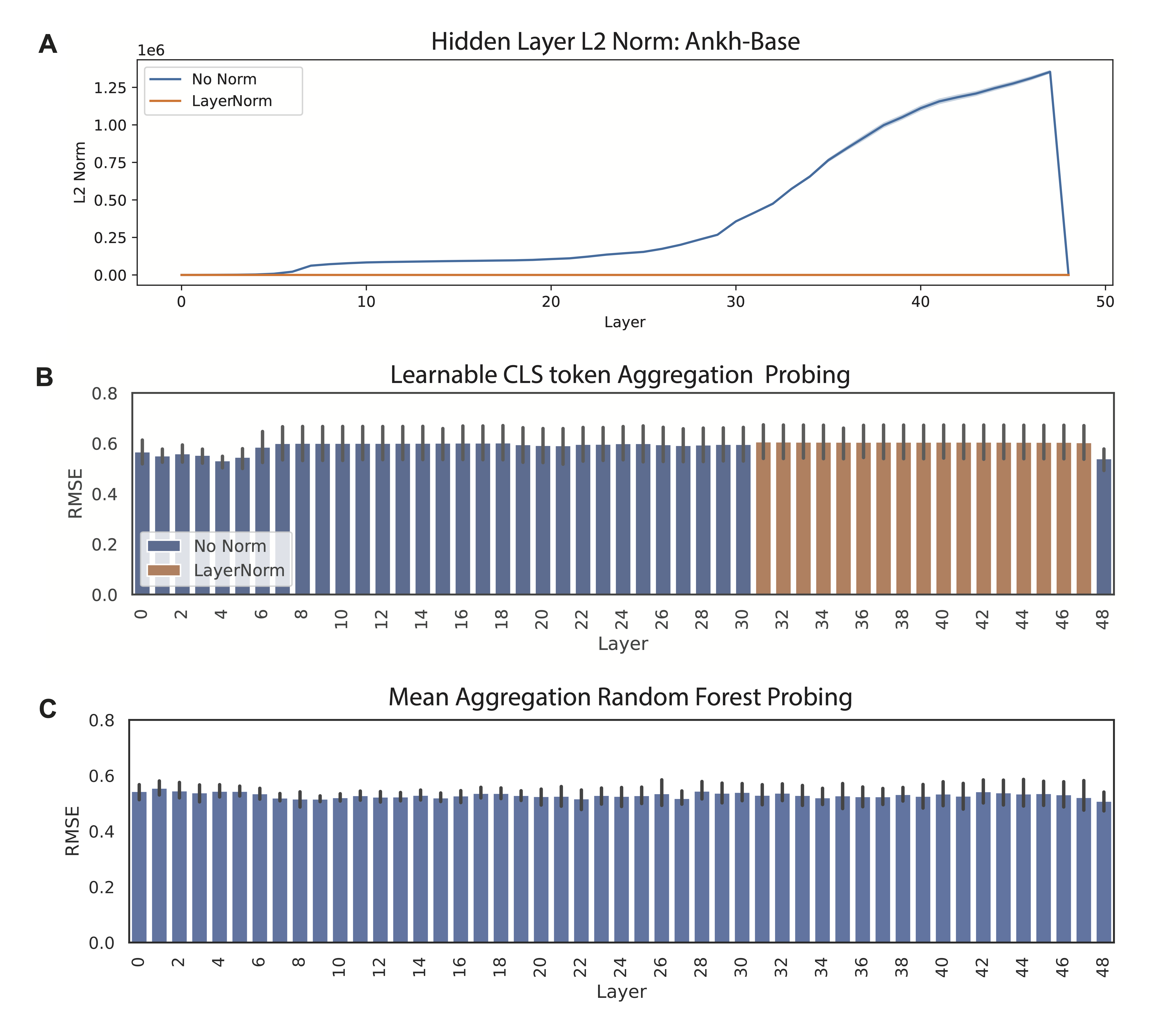}
    \caption{(A) L2 norm per layer before and after LayerNorm. (B) The mean validation RMSE for the $k_{cat}$ regression per layer using the ViT-Tiny CLS learnable aggregation method for the Ankh-base PLM embeddings. The error bar represents the standard deviation from the 5-fold cross-validation split. Note that Layers 31 to 47 are probed after LayerNorm, as the raw embeddings are too large and induce numerical instability during model training. (C) The mean validation RMSE for the $k_{cat}$ regression per layer using random forest probing on mean aggregated Ankh-base PLM embeddings.}
    \label{fig:enter-label}
\end{figure}

Additionally, we found that random forest probing of the mean aggregated embeddings was more stable and performant across the encoder layers (\textbf{Figure 2C}). While some layers' embeddings provided equivalent performance for the $k_{cat}$ regression task, they did not provide any benefit beyond the more widely used last layer embeddings.

\section{Discussion and Future Directions}

In this study, we sought to assess the state of the field for leveraging PLMs to predict $k_{cat}$ from enzyme sequence on a diverse set of ADK orthologs. We found that all PLM-based probing models outperformed existing larger $k_{cat}$ prediction models trained on public databases curated across multiple studies. Differences in performance among various PLM embedding configurations were modest, but we noted several trends. First, learnable aggregation functions for full-length embeddings only marginally outperform commonly used pooling strategies for constructing fixed-length embeddings. Second, using the Ankh-base model, we showed that additional MLM fine-tuning on related protein sequences (ADKs or EC 2.7) does not improve downstream $k_{cat}$ regression. Furthermore, directly fine-tuning using LoRA appears to be over-parametrized in the low-data regime and hinders model generalization on held-out sequences. Lastly, using embeddings from earlier layers in the Ankh-base encoder at best achieves comparable performance to using the final layer, but can introduce numerical instability. 

As the field of protein engineering looks to apply advances in ML and AI to accelerate the optimization of enzymatic activity, it is imperative to understand how to predict these parameters most accurately. 
Our results demonstrate the utility of high-throughput enzymology datasets in training models to predict catalytic turnover. They also underscore the critical need for additional data generation to improve models and understand how to generalize across proteins. 
Furthermore, as new protein design methods are capable of generating highly diverse sequences, our datasets and predictive models must explore broad regions of protein sequence space beyond single variant effects. 
Looking forward, we envision a synergistic scale-up of high-throughput enzymology assays and advancements in model architectures to enable efficient protein function optimization.

\begin{ack}
We would like to thank Dr. Tony Capra and Dr. Adam Yala for scientific feedback.
\end{ack}

\clearpage 

\printbibliography

\clearpage

\appendix

\section{Cross-Fold Validation: ADK Sequence Splits Stratified by Lid Type}
\begin{table}[h]
\centering
\caption{Organism Names for Orthologous ADK Splits (Lid type stratified)}
\begin{tabular}{lllll}
\hline

Split 1   & Split 2 & Split 3   & Split 4   & Split 5  \\
\hline
A. salm & A. mari & A. arab & A. sera & F. stag \\
A. deha & A. pleu & A. flav & A. muco & H. halo \\
A. caul & A. subb & A. ferr & A. guan & K. avic \\
B. hens & A. hain & A. bayl & A. medi & L. eryt \\
B. bact & B. cere & A. ehrl & B. aggr & M. hyor \\
C. trac & B. lich & A. orem & C. arse & M. mobi \\
C. urea & B. subt & A. fulg & D. alke & M. pneu \\
D. dese & B. quin & C. subt & D. acet & O. mess \\
E. minu & B. coag & C. japo & D. ther & P. bovi \\
E. tasm & B. petr & C. parv & E. pohl & P. umbo \\
F. nodo & B. pilo & C. apic & E. cate & P. medi \\
G. diaz & B. vulp & D. aura & E. coli & P. hydr \\
K. olea & C. ochr & D. prot & F. acid & P. maca \\
L. inte & C. pine & D. hafn & F. isla & P. mari \\
M. silv & C. botu & D. atla & G. aqua & P. debo \\
M. myco & E. faec & D. ther & G. puni & P. rhiz \\
P. magn & G. kaus & D. alim & H. acin & P. meth \\
P. mobi & G. stea & F. rode & H. rode & R. fulv \\
P. prof & G. ther & F. inte & H. ther & R. floc \\
P. chlo & G. dalt & F. hyda & L. mult & S. rube \\
P. atla & H. mode & H. mari & L. cris & S. paci \\
P. lett & H. somn & H. sali & M. naut & S. muco \\
P. furi & L. delb & J. mari & M. pauc & S. pleo \\
S. frig & L. lact & M. arvo & M. floc & S. subs \\
S. hali & L. pneu & M. caps & M. phyc & S. ther \\
S. cell & L. reut & M. ther & M. frap & S. alba \\
S. alas & L. spha & N. ther & M. choc & S. subr \\
T. sibi & M. burt & P. ingr & M. viri & S. elon \\
T. afri & M. acet & S. aren & M. argi & T. atla \\
T. mela & N. gono & S. loih & N. ulva & T. comm \\
T. mari & P. naph & S. pneu & O. oeni & T. rose \\
T. neap & P. ento & S. coel & P. hart & T. puti \\
V. chol & P. syri & S. ther & P. exov & V. semi \\
V. para & S. ther & S. wolf & R. mass & V. apor \\
X. camp & T. lovl & T. yell & R. radi & V. dokd \\
\hline
\end{tabular}
\end{table}

\section{Average L2 Norm of Final Layer PLM Embeddings}

\begin{figure}[h]
    \centering
    \includegraphics[width=0.7\linewidth]{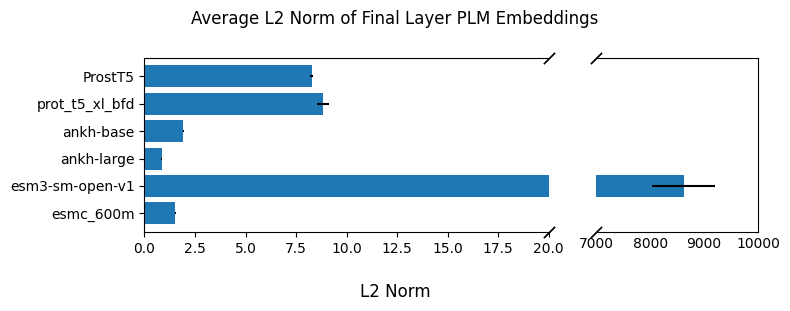}
    \caption{PLM amino acid embedding L2 norms. Error bars represent std. dev.}
    \label{fig:enter-label}
\end{figure}

\section{Reproducibility of MLP Results Across Initialization Random Seeds}

To assess the reproducibility of the MLP probing results, we used two random seeds and plotted the consistency of the RMSE on the regression task. Models trained on ESM3 are not plotted here due to their large error values, likely stemming from large-magnitude embeddings which induced numerical instability during training.

\begin{figure}[h]
    \centering
    \includegraphics[width=0.7\linewidth]{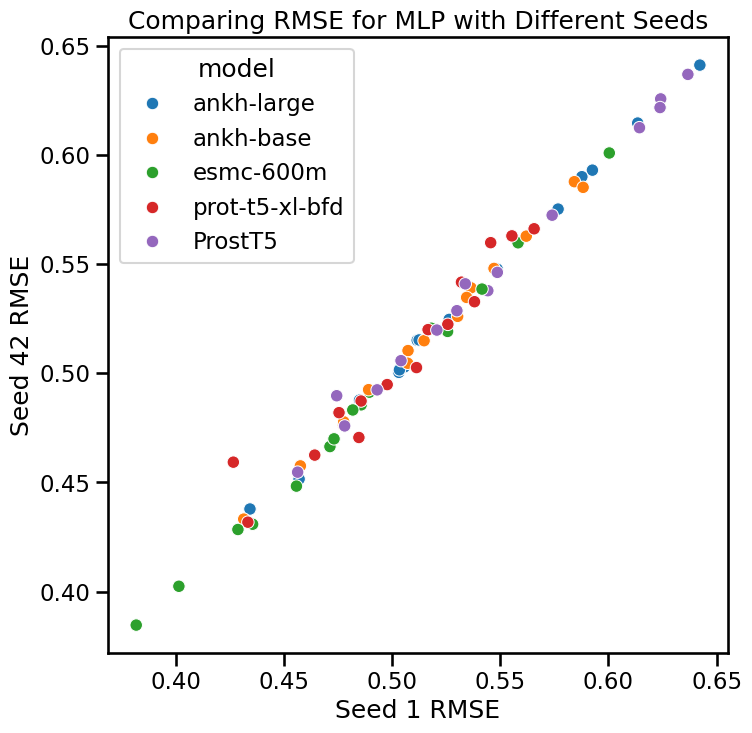}
    \caption{Comparison of random seed initialization for MLP Training with each point representing a validation fold for each of the models probed with an MLP.}
    \label{fig:enter-label}
\end{figure}

\section{Comparing probing of PLM embeddings to baselines}
Although this study focuses on experiments with PLM embeddings, we also evaluated baselines that do not rely on them. The first baseline we assessed was randomly initialized orthogonal embeddings for each amino acid matching the 768 dimensionality of the Ankh-base embeddings. These random embeddings sought to act as a pseudo-one-hot-encoding method. We also generated amino acid embeddings with cosine distances equal to the BLOSUM62 matrix via optimization with the Adam optimizer. Using random forest probing we found that the Ankh-base PLM embeddings performed superior by all metrics for the downstream tasks compared to the baseline embedding methods.

\begin{table}[h]
\centering
\caption{Comparing the random forest probing performance for Ankh-base embeddings with randomly distributed embeddings of the same dimension and embeddings generated to have cosine distances equal to the BLOSUM62 matrix}
\begin{tabular}{lllll}
\toprule
 & Embeddings & RMSE ($\downarrow$) & PCC ($\uparrow$) & $R^2$ ($\uparrow$) \\
 \midrule
 & Ankh-Base Zero Shot (Mean Agg.) & \textbf{0.509 $\pm$ 0.043} & \textbf{0.574 $\pm$ 0.142} & \textbf{0.312 $\pm$ 0.159} \\
  \cline{1-5}
 & Pseudo BLOSUM62 (Mean Agg.)   & 0.539 $\pm$ 0.050 & 0.515 $\pm$ 0.133 & 0.233 $\pm$ 0.129  \\
 \cline{1-5}
 & Pseudo one-hot (Mean Agg.)  & 0.553 $\pm$ 0.030 & 0.464 $\pm$ 0.198 & 0.178 $\pm$ 0.205      \\  
 \cline{1-5}
 \bottomrule
\end{tabular}
\end{table}

\end{document}